%% file: Wunderer_2019_asset_correlation_estimation_for_inhomogeneous_exposure_pools.ltx
\documentclass[11pt]{article} 

\usepackage[utf8]{inputenc} 

\usepackage{geometry} 
\usepackage{graphicx} 

\usepackage{booktabs} 
\usepackage{array} 
\usepackage{paralist} 
\usepackage{verbatim} 

\usepackage{fancyhdr} 
\usepackage{sectsty}
\allsectionsfont{\sffamily\mdseries\upshape} 

\usepackage[nottoc,notlof,notlot]{tocbibind} 
\usepackage[titles,subfigure]{tocloft} 

\input{customisation.tex}
\usepackage{placeins}
\usepackage{caption}
\usepackage{natbib}
\usepackage{subcaption}
\usepackage{float}

\title{Asset correlation estimation for inhomogeneous exposure pools}
\author{Christoph Wunderer\footnote{Please direct correspondence to christoph.wunderer@s-rating-risikosysteme.de, Sparkassen Rating- und Risikosysteme GmbH, Leipziger Straße 51, 10117 Berlin; the opinions expressed here are those of the author, and do not reflect the views of Sparkassen Rating and Risikosysteme GmbH or its staff. Christoph Wunderer would like to thank Hubert Eckelmann and two anonymous referees for helpful comments during the preparation of the manuscript.}}

\begin{document}

\newcolumntype{C}{>{\centering\arraybackslash}p{2em}}

\maketitle
\thispagestyle{fancy}

\begin{abstract}
 \input{paper_abstract.tex}
\end{abstract}
\section{Introduction}
\input{paper_introduction.tex}

\section{Inhomogeneous exposure pools}\label{theo}
 \input{paper_theo.tex}

\section{A numerical study}\label{num}
 \input{paper_study.tex}

\section{Empirical evidence}\label{lit}
 \input{paper_literature.tex}

\section{Discussion}\label{disc}
 \input{paper_discussion.tex}

\bibliographystyle{apa}
\bibliography{./credit} 

\pagebreak

\section{Appendix}\label{appen}
 \input{paper_appendix.tex}


\end{document}

%% file: customisation.tex
\usepackage{url}
\usepackage{amssymb,amsmath,amsthm, amsfonts}

\graphicspath{ {.} }

\newcommand{\cov}{\mathrm{cov}}
\newcommand{\var}{\mathrm{var}}

\newcommand{\E}{\mathrm{E}}
\newcommand{\Prob}{\mathrm{P}}
\newcommand{\DRb}{\overline{D\!R}}
\newcommand{\DR}{D\!R}
\newcommand{\pbar}{\bar{p}}
\newcommand{\rhoabar}{\bar{\rho}^A}
\newcommand{\rhoatilde}{\tilde{\rho}^A}
\newcommand{\rhoahat}{\hat{\rho}^A}
\newcommand{\rhoa}{\rho^A}

\newcommand{\refe}[1]{(\ref{#1})}

%% file: paper_abstract.tex
\noindent A possible data source for the estimation of asset correlations is default time series. This study investigates the systematic error that is made if the exposure pool underlying a default time series is assumed to be homogeneous when in reality it is not. We find that the asset correlation will always be underestimated if homogeneity with respect to the probability of default (PD) is wrongly assumed, and the error is the larger the more spread out the PD is within the exposure pool. If the exposure pool is inhomogeneous with respect to the asset correlation itself then the error may be going in both directions, but for most PD- and asset correlation ranges relevant in practice the asset correlation is systematically underestimated. Both effects stack up and the error tends to become even larger if in addition a negative correlation between asset correlation and PD is assumed, which is plausible in many circumstances and consistent with the Basel RWA formula. It is argued that the generic inhomogeneity effect described is one of the reasons why asset correlations measured from default data tend to be lower than asset correlations derived from asset value data.

\smallskip
\noindent {\bf Keywords:} asset correlation, default time series, credit portfolio risk, downward bias, inhomogeneity effect

%% file: paper_introduction.tex
Most simulation based portfolio models in use today do not simulate default events directly but rather employ a structural model of default where a continuous variable is simulated, sometimes termed \lq\lq{}creditworthiness index\rq\rq{} (\cite{Bluhm_et_al_2007}) or \lq\lq{}ability-to-pay variable\rq\rq{} (\cite{Kalkbrener_et_al_2009}). The default of an exposure is recorded if its creditworthiness index falls below a certain threshold linked to the PD of that exposure. The correlations between different cre\-ditworthiness indices are called \lq\lq{}asset correlations\rq\rq{}. If a structural model of default is employed then asset correlations are crucial to parametrize it and their estimation is important with direct impact on modeling results.

This paper is concerned with the estimation of asset correlations from default time series, a method that is particularly important when the modeled exposures do not belong to listed companies and alternative estimation techniques based on market data are not available. Several authors have observed that asset correlations estimated from default time series tend to be lower than asset correlations estimated from stock price time series (\cite{Frye_2008},  \cite{Duellmann_et_al_2008}, \cite{Kalkbrener_et_al_2009}, \cite{Chernih_et_al_2010}). As a possible explanation \cite{Frye_2008} identifies the approximate nature of the structural model of default that links asset correlations to default data. Another very valid explanation is that the estimators commonly used for the estimation of asset correlations from default time series have a downward bias leading to low asset correlation estimates (\cite{Gordy_et_al_2010}, \cite{Duellmann_et_al_2008}, \cite{Frei_et_al_2018}). As the bias of many estimators depends more on the length of the time series and less on the number of exposures taken into account for the time series this bias tends to have a noticeable size in practice (\cite{Meyer_2009}). While there are many questions to be investigated with respect to this bias and how it potentially could be corrected, this paper is not following this route but describes an entirely different mechanism that to our knowledge has not been discussed in the literature in detail yet: if a pool of exposures is assumed to be homogeneous with respect to PD and/or asset correlation but in fact is not, then the asset correlation of that exposure pool is measured too low in a systematic way. As any real-life exposure pool will be inhomogeneous to some degree, this mechanism provides an additional explanation for the puzzle why asset correlations derived from default rates tend to be lower than asset correlations derived directly from asset value data where such an inhomogeneity effect is not relevant.

Understanding the inhomogeneity effect has direct implications for the practitioner who should ensure that asset correlations are only estimated for homogeneous exposure pools or that estimators are used that explicitly take the inhomogeneity of the studied exposure pool into account. As there will always be some residual inhomogeneity present in the chosen setup it is prudent to assume that in addition to the already well known downward bias of asset correlation estimators there is another downward effect due to the residual inhomogeneity of the studied exposure pool.

In section \ref{theo} the underlying theory is presented, much of which is standard and can be found for instance in  \cite{Bluhm_et_al_2010},  \cite{McNeil_et_al_2015}  and \cite{Lucas_1995}. As main result of the section we learn how the variance of a default rate belonging to an inhomogeneous exposure pool can be linked to the PD and asset correlation distribution within that exposure pool.

In section \ref{num} a fairly general setup is described that allows us to study numerically the systematic underestimation of asset correlations under a great variety of parameter regimes. In particular we study the dependence of the inhomogeneity effect on the location and shape of the PD- as well as the asset correlation distribution within a given exposure pool. We also study the effect of correlated PD- and asset correlation profiles.

In  section \ref{lit} we point to the literature for some empirical evidence and in section \ref{disc} we discuss the results obtained.

%% file: paper_theo.tex
We consider a pool of $n$ exposures distributed across $K\times L$ sub-pools with $n_{kl}$ exposures in the sub-pool labeled $(k, l)$, such that $\sum_{k=1}^K\sum_{l=1}^L n_{kl} = n$.  It is assumed that all sub-pools are homogeneous with respect to PD and asset correlation, that for a given $k$ all sub-pools $(k, 1\le l \le L)$ share the same PD, which we will call $p_k$, and that for a given $l$ all sub-pools $(1\le k \le K, l)$ share the same asset correlation, which we will call $\rhoa_l$. The default event and the default indicator random variable are defined as follows:
\begin{eqnarray}
D^i_{kl}&=&\{ \mbox{exposure }  i \mbox{ in sub-pool } (k,l) \mbox{ defaults} \}\,\nonumber  \\ 
1_{D^i_{kl}} &=&\left\{\begin{array}{ll} 1 & \mbox{if exposure }  i \mbox{ in sub-pool } (k,l) \mbox{ defaults }  \\ 0 & \mbox{otherwise}\end{array} \right.\nonumber
\end{eqnarray}
Throughout the paper it is understood that the horizon for the observation of a default event is always the same, typically one year. The PD  of an exposure $i$ is the expected value of the default indicator random variable
\begin{equation}
p_k=\Prob\left( D^i_{kl} \right) =\E\left( 1_{D^i_{kl}} \right)\nonumber
\end{equation}
For each exposure $i$ in sub-pool $(k,l)$ we assume that its default event can be fully described via a structural model with one latent variable $z$ shared by all exposures and a creditworthiness index $y^i_{kl}$ that is specific to the exposure considered
\begin{equation}
y^i_{kl}=\sqrt{\rhoa_{l}}z+\sqrt{1-\rhoa_{l}}\epsilon^i_{kl}\label{structural}
\end{equation}
Here $z, y^i_{kl} , \epsilon^i_{kl}\sim N(0,1)$ are standard normally distributed, and $z$ as well as all the idiosyncratic factors $\epsilon^i_{kl}$ are pairwise independent. The constant $\rhoa_{l}$ is the asset correlation and the correlation of two different creditworthiness indices $y^i_{kl}$ and $y^j_{st}$ is given by $\sqrt{\rhoa_{l}\rhoa_{t}}$. In addition we assume that two creditworthiness indices have a joint bivariate normal distribution. See e.g. \cite{Bluhm_et_al_2010} for some background on this widely used modeling approach. The basic assumption of the structural model of default is
\begin{equation}
D^i_{kl} \Leftrightarrow y^i_{kl} < c_{k} \ , \quad \mbox{where } c_k=\Phi^{-1}\left(  p_k \right)\nonumber
\end{equation}
For $i\ne j$, exposure $i$ belonging to sub-pool $(k,l)$ and exposure $j$ belonging to sub-pool $(s,t)$ we have:
\begin{eqnarray}
\cov \left(1_{D^i_{kl}}  , 1_{D^j_{st}}   \right) &=& \E \left(1_{D^i_{kl}}  1_{D^j_{st}} \right) - \E \left(1_{D^i_{kl}}  \right)\E \left(  1_{D^j_{st}}  \right) =\Phi_2 \left(c_{k} , c_{s} , \sqrt{\rhoa_l\rhoa_t}\right) - p_k p_s\nonumber \\
\var \left(1_{D^i_{kl}}    \right) &=& \E \left(1_{D^i_{kl}}^2 \right) - \E \left(1_{D^i_{kl}}  \right)^2 = p_k -p_k^2\label{covid}
\end{eqnarray}
Instead of single exposures we now consider an entire sub-pool $(k,l)$ with $n_{kl}$ exposures. The default rate random variable $\DR_{kl}$ is obtained by averaging $n_{kl}$ default indicators:
\begin{equation}
\DR_{kl}=\sum_{i=1}^{n_{kl}}\frac{1_{D^i_{kl}}}{n_{kl}}\quad , \quad E(\DR_{kl})=p_k\nonumber\label{edr}
\end{equation}
Using \refe{covid} it follows for two different sub-pools $(k,l)$ and $(s,t)$ that
\begin{eqnarray}
\cov(\DR_{kl}, \DR_{st})&=& \frac{1}{n_{kl}n_{st}}\sum_{i=1}^{n_{kl}}\sum_{j=1}^{n_{st}}\cov  \left(1_{D^i_{kl}}  , 1_{D^j_{st}}   \right)= \nonumber \\
&=&\Phi_2 \left(c_{k} , c_{s} , \sqrt{\rhoa_l\rhoa_t}\right) - p_k p_s\label{covardrin}\\ 
\var(\DR_{kl})&= &\frac{1}{n_{kl}^2}\left( \sum_{i\neq j}^{n_{kl}}\cov  \left(1_{D^i_{kl}}  , 1_{D^j_{kl}}   \right) +  \sum_{i=1}^{n_{kl}}\var \left(1_{D^i_{kl}}    \right) \right)= \nonumber\\
&=&\Phi_2 \left(c_{k} , c_{k} , \rhoa_l \right) - p_k^2 +\frac{p_k -\Phi_2\left( c_k , c_k , \rhoa_l\right)}{n_{kl}}\label{covardrout}
\end{eqnarray}
Combining the $K\times L$ sub-pools we can define the overall default rate $\DRb$ as
\begin{eqnarray}\label{combined2}
\DRb=\sum_{k=1}^K\sum_{l=1}^L\frac{n_{kl} \DR_{kl}}{n}\nonumber
\end{eqnarray}
Defining the mean PD as $\pbar=\frac{1}{n}\sum_{k=1}^{K}\sum_{l=1}^{L} p_k n_{kl}=\E (\DRb)$ and using equations \refe{covardrin} and \refe{covardrout} we obtain
\begin{equation}
\var(\DRb)=\sum_{k,s=1}^K \sum_{l,t=1}^L \frac{n_{kl}}{n} \frac{n_{st}}{n} \Phi_2 \left( c_k ,c_s, \sqrt{\rhoa_l\rhoa_t} \right) - \pbar^2 +\frac{\pbar-\sum_{k=1}^K \sum_{l=1}^L \frac{n_{kl}}{n} \Phi_2 \left( c_k , c_k , \rhoa_l\right)}{n} \label{multibucket3}
\end{equation}
Equation \refe{multibucket3} allows us to calculate the variance of the default rate of an inhomogeneous exposure pool. Before we proceed to assess the effect of the assumption of homogeneity, we consider two special cases of equation \refe{multibucket3}.

First we consider $K=L=1$, i.e. we assume the exposure pool to be homogeneous with respect to PD and asset correlation and set $p=p_{1}$, $\rhoa=\rhoa_1$, $c=c_1$ to obtain
\begin{equation}
\var(\DRb)= \Phi_2 \left(c , c , \rhoa\right) - p^2 +\frac{p -\Phi_2\left( c , c , \rhoa\right)}{n}\label{singlebucket}
\end{equation}
Equation \refe{singlebucket} is equivalent to equation \refe{covardrout}. Note that $\var (\DRb )$ increases strictly with $\rhoa$, such that $\rhoa\in [ 0,1 ]$ can be backed out numerically if $\var(\DRb)$, $p$ and $n$ are known and if $\var (\DRb )$ lies between the following bounds:
\begin{equation}
\frac{(1-p)p}{n}\leq \var (\DRb) \leq (1-p)p\label{varbound}
\end{equation}
Equation (\ref{singlebucket}) provides a straightforward method for estimating the asset correlation from an observed default time series: first $p$ is estimated as the average observed default rate and then $\var (\DRb )$ is estimated as the sample variance of the periodical default rate observations. Finally $\rhoa$ is backed out from (\ref{singlebucket}) as described. Note, however, that this estimator, often referred to as \lq\lq{}method of moments\rq\rq{} estimator, is biased:  the estimated asset correlation tends to be underestimated and the shorter the available time series of default rate observations is, the larger this underestimation will be on average. This observation is linked to the fact that the function $\var (\DRb) \longrightarrow \rhoa\left(\var (\DRb )\right)$ is concave (see \cite{Gordy_et_al_2010}). This estimation bias, however, is not the focus of this study and also plays no role for its main conclusions.

We also consider $L=1$, $K>1$ as another special case of equation \refe{multibucket3}, i.e. we assume the pool of exposures to be homogeneous with respect to the asset correlation, but not the PD. After setting  $\rhoa=\rhoa_1$ and $n_k=n_{k1}$ we obtain
\begin{equation}
\var(\DRb)= \sum_{k,s=1}^K \frac{n_k}{n} \frac{n_s}{n} \Phi_2 ( c_k ,c_s , \rhoa ) - \pbar^2+\frac{\pbar-\sum_{k=1}^K \frac{n_k}{n} \Phi_2 ( c_k , c_k , \rhoa)}{n}\label{multibucket}
\end{equation}
Rearranging terms and noting that  $\Phi_2(c_k, c_s, \cdot)$ is strictly increasing we can see that $\var(\DRb)$ is a strictly increasing function of $\rhoa$ such that  $\rhoa$ can be backed out if $n_k , p_k$ and $\var(\DRb)$ are given and $\var(\DRb)$ lies between certain bounds. Thus an estimator for $\rhoa$ can be defined that is applicable to an inhomogeneous pool of exposures but  the properties of this estimator and how it compares to the estimator MLE3 of \cite{Gordy_et_al_2010} that serves the same purpose are not within the scope of this paper.

%% file: paper_study.tex
\subsection{Setup}
In order to apply equation \refe{multibucket3} to an exposure pool first a certain constellation of homogeneous sub-pools is chosen, which is characterized by the following parameters, the combination of which we will call {\em exposure constellation}:
\begin{itemize}
\item $K$: number of PD buckets used
\item $L$: number of asset correlation buckets used
\item $\rhoa_l , 1\le l \le L$: different $\rhoa$ values for the $L$ asset correlation buckets
\item $p_k , 1\le k \le K$: different PD values for the $K$ PD buckets
\item $n_{kl} , 1\le k \le K , 1\le l \le L$: number of exposures in each of the $L\times K$ exposure pools
\end{itemize}
Once an exposure constellation is chosen the overall asset correlation is calculated in two ways. Once directly by averaging the asset correlation across the exposure pools defined in the exposure constellation leading to $\rhoabar$, and once by using the information contained in the exposure constellation to calculate the variance of the overall default rate $\DRb$ via equation (\ref{multibucket3}) and then backing out  $\rhoatilde$ from that variance of the default rate via equation (\ref{singlebucket}), assuming the underlying pool of exposures was homogeneous. $\rhoatilde$ is the asset correlation that would be measured\footnote{we write here, and throughout the paper, \lq\lq{}measured\rq\rq{} to stress the error potential for the practitioner, but our approach derives the asset correlation from the analytical properties of the exposure constellation and no estimation error is included in the analysis.}  under the (wrong) assumption that the exposure constellation was in fact homogeneous with respect to PD and asset correlation. The discrepancy between $\rhoatilde$ and the average asset correlation $\rhoabar$  is a measure of the error that is made by making the assumption of homogeneity. In the results sections we look at the ratio $\rhoa\%=\rhoatilde / \rhoabar$ to characterize this discrepancy.

One expects $\rhoa\%$ to depend on the properties of the exposure constellation, the degree of inhomogeneity being one of them. The exposure constellation is characterized by several parameters, the combination of which we call {\em input configuration}:
\begin{itemize}
\item $n=\sum_{k=1}^{K}\sum_{l=1}^{L} n_{kl}$: the overall number of exposures
\item $\pbar=\frac{1}{n}\sum_{k=1}^{K}\sum_{l=1}^{L} n_{kl}p_k$: the mean PD
\item $\sigma (p) =\sqrt{\frac{1}{n}\sum_{k=1}^{K}\sum_{l=1}^{L}n_{kl}(p_k - \pbar)^2}$: the standard deviation of the PD profile
\item $\rhoabar=\frac{1}{n}\sum_{k=1}^{K}\sum_{l=1}^{L} n_{kl}\rhoa_l$: the average asset correlation
\item $\sigma (\rhoa) =\sqrt{\frac{1}{n}\sum_{k=1}^{K}\sum_{l=1}^{L}n_{kl}(\rhoa_l - \rhoabar)^2}$: the standard deviation of the $\rhoa$ profile
\item $\tau(p,\rhoa)$: Kendall\rq{}s rank correlation coefficient of the asset correlation and PD profiles, see \cite{Lindskog_2003} for details on its relation to the linear correlation coefficient. For its calculation we use the formula for tau-b, a variant of Kendall\rq{}s tau for discrete data that includes a correction for ties, see \cite{SAS_1999}:
\end{itemize}
\begin{eqnarray}
&& \tau(p,\rhoa)=\frac{\sum_{k=1}^K\sum_{l=1}^Ln_{kl}\left(A_{kl}-B_{kl}\right)}{\sqrt{D_r D_c}} , \mbox{where}\nonumber\\
&&A_{kl}=\sum_{k<i}\sum_{l<j}n_{ij}+\sum_{k>i}\sum_{l>j}n_{ij}, \quad B_{kl}=\sum_{k<i}\sum_{l>j}n_{ij}+\sum_{k>i}\sum_{l<j}n_{ij}\nonumber\\
&&D_r=n^2-\sum_{k=1}^{K}\left( \sum_{l=1}^{L} n_{kl}\right) ^2, \quad D_c=n^2-\sum_{l=1}^{L}\left( \sum_{k=1}^{K} n_{kl}\right) ^2\nonumber
\end{eqnarray}
Given an exposure constellation, we can diagnose the corresponding input con\-fi\-guration, but in order to study the dependence of $\rhoa\%$ on a given input configuration we need to derive an exposure constellation that corresponds to it. In order to do so we need to choose the number of buckets $K$ and $L$ according to our desire for accuracy and make some distributional assumptions to obtain the joint PD- and asset correlation profile $\{p_k , \rhoa_l , n_{kl}\}$. Technical details of this construction based on the Gaussian copula and beta-distributed marginals can be found in the appendix. 

\subsection{Inhomogeneous profile of PDs}\label{inh1}

\begin{figure}[!htbp]
\centering
\includegraphics[width=0.8\linewidth]{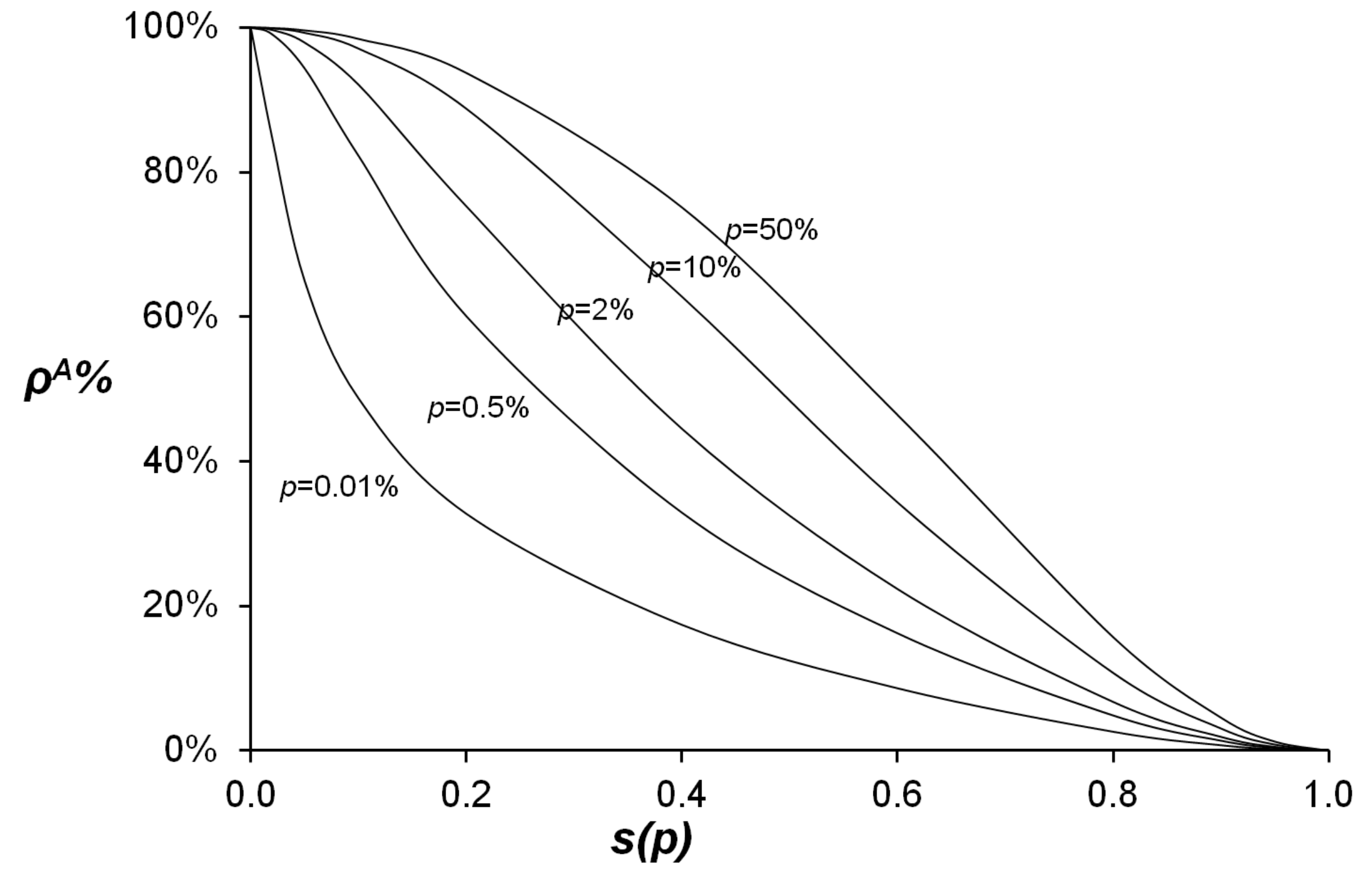}
\caption{Measured asset correlation as percentage of actual asset correlation, using $n=10^9$, $K=1000$, $L=1$ , $\rhoa=12\%$.}
 \label{figure_onion_02}
\end{figure}

For the first series of results, we assume the asset correlation $\rhoa$ to be homogeneous within the given pool of exposures, i.e. $L=1$, but we assume a distribution of PDs. For the results in this section we choose $K=1000$. We furthermore choose $n=10^9$ to be very large and $\rhoa=12\%$, which is a value that lies within the range that is relevant in practice. Figure \ref{figure_onion_02} shows that for different values of $\pbar$ and $s(p)=\frac{\sigma(p)}{\sqrt{\pbar(1-\pbar)}}$ we always have $\rhoa\%<100\%$, i.e. the measured asset correlation is always reduced if a homogeneous exposure pool is assumed. We use $s(p)$ for characterizing the width of the PD distribution instead of just $\sigma(p)$ in order to make standard deviations for different values of $\pbar$ comparable. 

If $s(p)=0$, i.e. if the PD is constant throughout the pool, there is no reduction of the asset correlation, as in this case the equations (\ref{singlebucket}) and (\ref{multibucket3}) are identical. $s(p)=1$ implies that the portfolio is split into two exposure pools with PDs $p_1=0$ and $p_2=100\%$. The overall default rate ceases to be random and $\var (\DRb) = 0$. Backing out the asset correlation from (\ref{singlebucket}) leads to $\rhoatilde=0$ in the limit $n\rightarrow \infty$. For $0<s(p)<1$ the backed out asset correlation is reduced and this reduction increases the larger $s(p)$ gets. 

Note that the picture presented in  Figure \ref{figure_onion_02} does not change significantly if we vary the asset correlation $\rhoa$.
 
\begin{figure}[!htbp]
\centering
\includegraphics[width=0.8\linewidth]{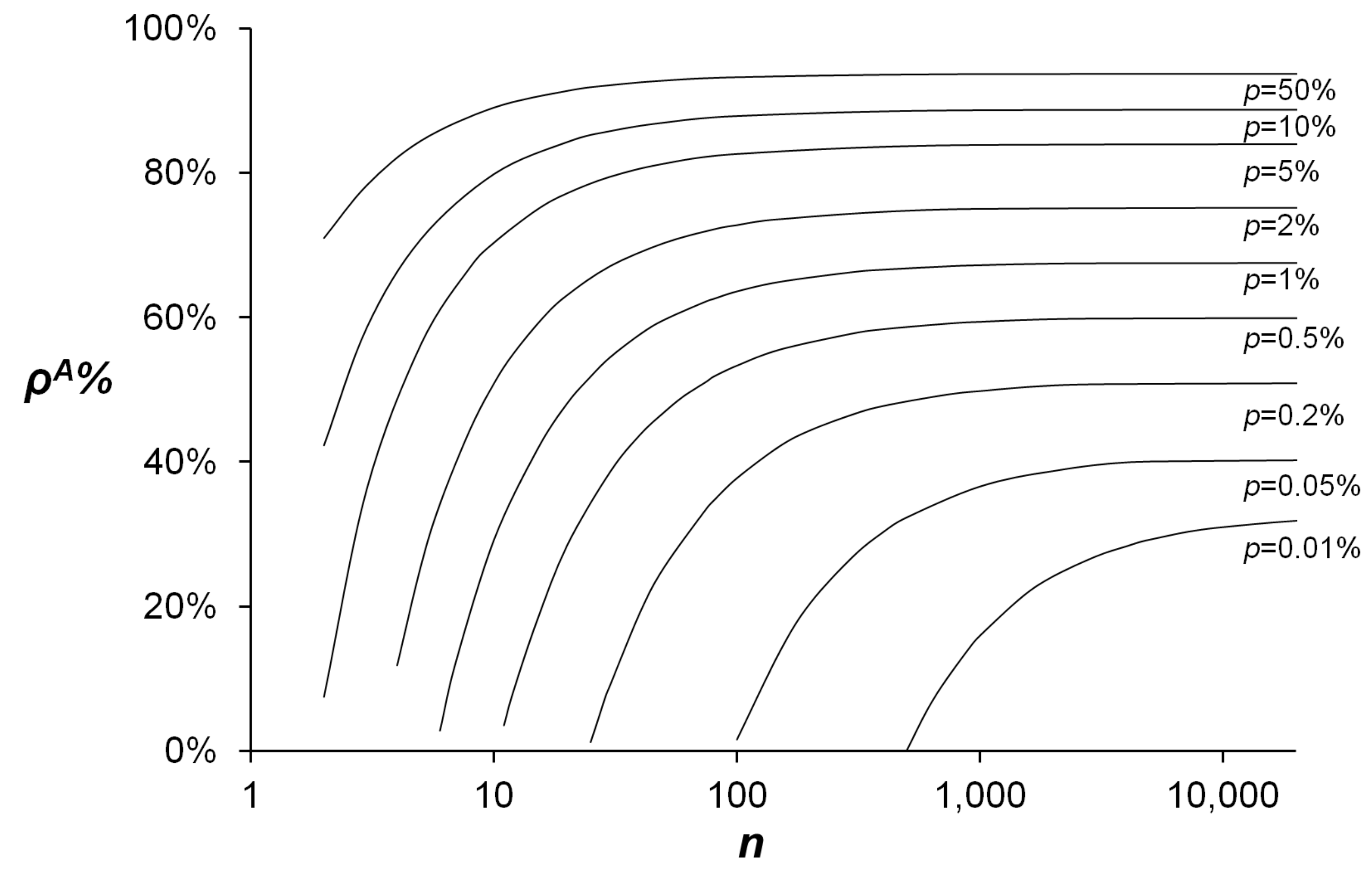}
\caption{Measured asset correlation as percentage of actual asset correlation, $K=1000$, $L=1$ , $\rhoa=12\%$ , $s(p) = 20\%$. }
 \label{figure_onion_06}
\end{figure}

For the results presented so far, we have set the number of exposures $n$ to $10^9$. For small $n$, however, the asset correlation reduction effect  is increased. If the PD is small then this increase sets in already for fairly high numbers of $n$, see figure \ref{figure_onion_06} as an example. Note, however, that a small $n$ only has an amplifying effect on the asset correlation reduction induced by $s(p)>0$, it does not have an effect on its own as for $s(p)=0$ the equations (\ref{singlebucket}) and  (\ref{multibucket}) are identical. Note also that for a small $\pbar$ there is a minimum $n$ below which $\var (\DRb )$ obtained from equation (\ref{multibucket}) lies outside the bounds given in (\ref{varbound}) such that equation (\ref{singlebucket}) cannot be inverted. If $n$ is small then the placement of a single exposure into a PD bucket can affect the mean PD of the exposure pool, which leads to the observation that $\pbar$ of the input configuration and $\pbar$ of the exposure constellation may differ by more than $1\%$, which is the tolerance used elsewhere in this paper. The relative difference in the cases studied for figure \ref{figure_onion_06}, however, has always been less than $20\%$.

\FloatBarrier
\subsection{Inhomogeneous profile of asset correlations}\label{inh2}

We now assume a constant PD within an exposure pool but allow the asset correlation to have a distribution around its average $\rhoabar$. We take $K=1$ and $L=1000$ and proceed as above, but see a quite different picture: the asset correlation is not always reduced if $s(\rhoa)$ increases. Whether there is a reduction depends on $\rhoabar$ as well as $p$ and $s(\rhoa)$.

\begin{figure}[!htbp]
\centering
\includegraphics[width=0.8\linewidth]{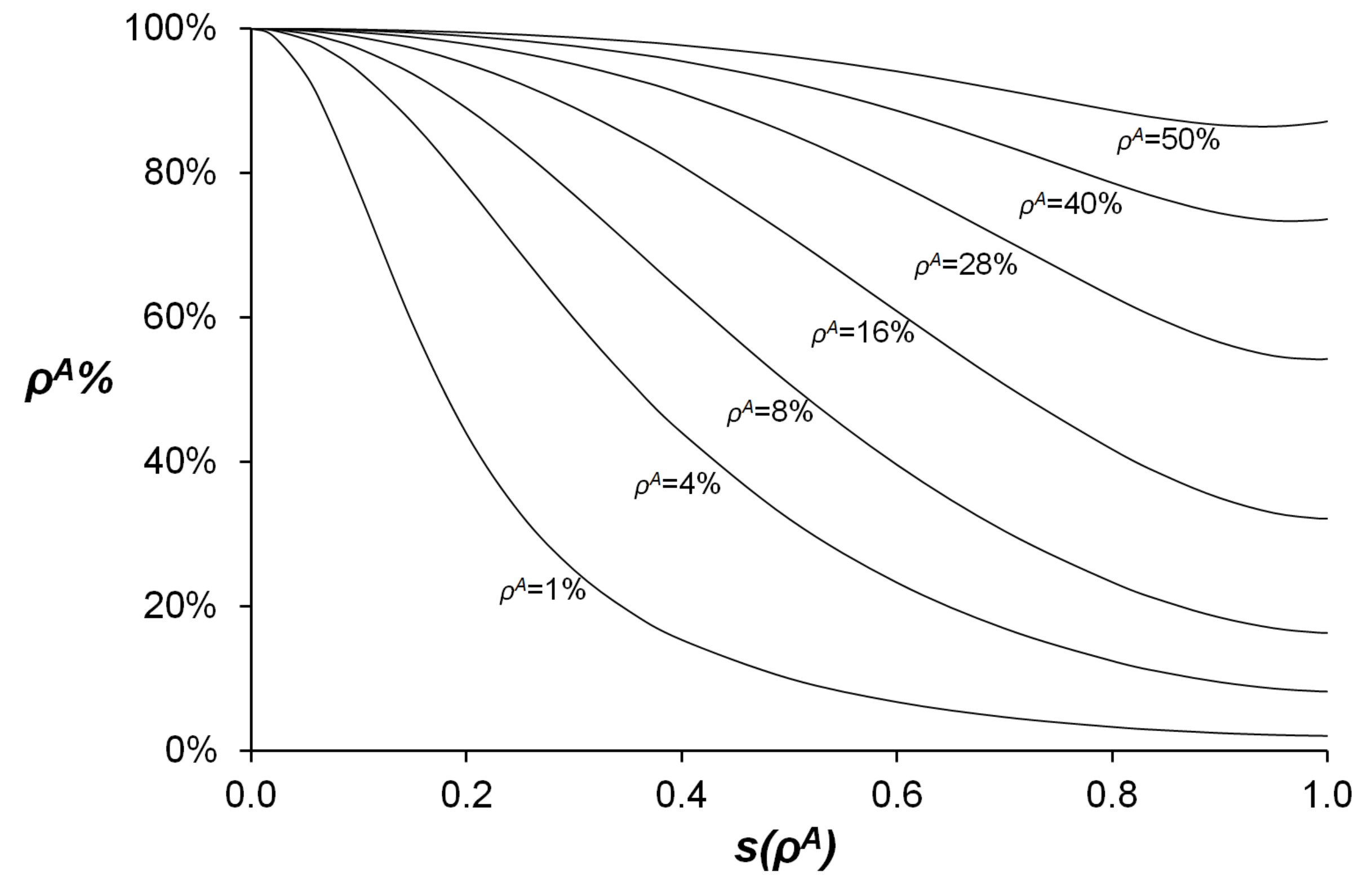}
\caption{Measured asset correlation as percentage of actual mean asset correlation, using $n=10^9$, $K=1$, $L=1000$ , $p=20\%$. }
 \label{figure_onion_07}
\end{figure}

\begin{figure}[!htbp]
\centering
\includegraphics[width=0.8\linewidth]{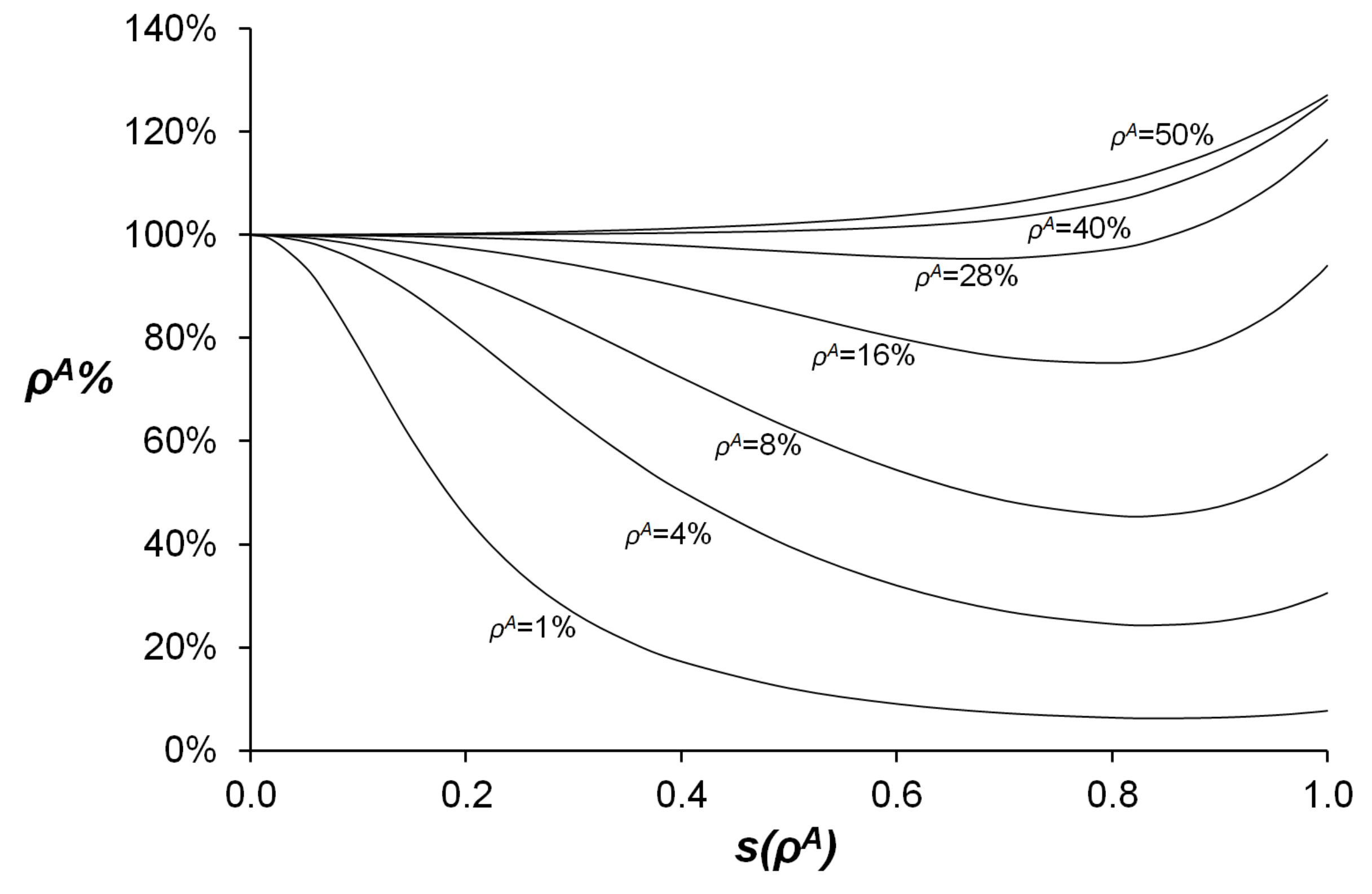}
\caption{Measured asset correlation as percentage of actual mean asset correlation, using $n=10^9$, $K=1$, $L=1000$ , $p=2\%$. }
 \label{figure_onion_08}
\end{figure}

\begin{figure}[!htbp]
\centering
\includegraphics[width=0.8\linewidth]{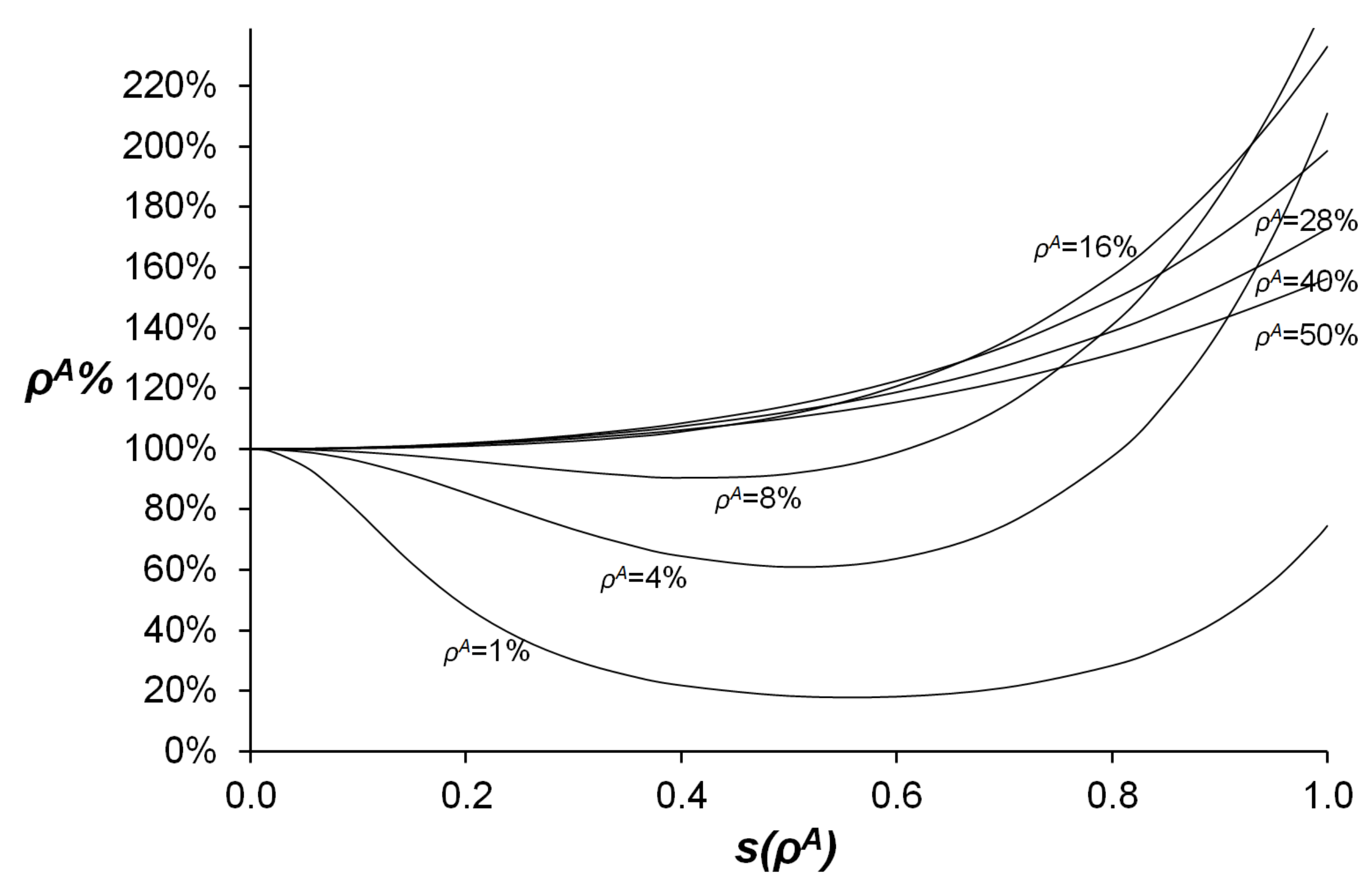}
\caption{Measured asset correlation as percentage of actual mean asset correlation, using $n=10^9$, $K=1$, $L=1000$ , $p=0.1\%$. }
 \label{figure_onion_09}
\end{figure}

The results presented in figures \ref{figure_onion_07} - \ref{figure_onion_09} suggest, however, that for parameter regimes very relevant in practice we can expect a reduction of the asset correlation. This is in particular the case if $p>1\%$ and $\rhoabar<20\%$ but also for lower PDs as long as  $\rhoabar$ is small. Only for very low PDs and fairly high $\rhoabar$ we expect an increase of the asset correlation, in particular if $s(\rhoa)$ is high as well. 

\FloatBarrier
\subsection{Inhomogeneous profiles of PD and asset correlation}\label{inh3}

Now we consider pools that are inhomogeneous with respect to both PD and asset correlation. As the calculations become more costly numerically we only use $K=200$ and $L=100$, but still achieve an accurate reflection of the input configurations we are interested in if we set $g=10^6$ and $ p_{\mathrm{mid}}=p_{\mathrm{median}}$ (for these technical parameters see the appendix). There are only important differences for $\pbar<0.1\%$, a PD range we therefore exclude from the analysis.

For a given input configuration we calculate $\rhoa\%$ and call it $\rhoa\%_{\rho p}$, then we set $s(p)=0\%$, calculate  $\rhoa\%$ and call it $\rhoa\%_{\rho}$ and finally we leave $s(p)$ unchanged, set  $s(\rhoa)=0\%$, calculate  $\rhoa\%$ and call it $\rhoa\%_{p}$. For all cases studied we find that the effects of inhomogeneous PD and inhomogeneous asset correlation stack up in an approximately multiplicative way:
\begin{equation}
 \rhoa\%_{\rho p} \approx \rhoa\%_\rho \rhoa\%_p\label{stacking}
\end{equation}
The tables \ref{table_zero1} and \ref{table_zero2} illustrate this relationship for $\rhoabar=4\%$ and  $\rhoabar=20\%$ as well as different values of $\pbar$.
\begin{table}[htbp]
\centering \footnotesize{
\begin{tabular}{@{}llllll@{}}
\cmidrule(l){3-6} 
\textbf{} & \textbf{} & \textbf{$\mathbf{\rhoa\%_\rho}$} &  \textbf{$\mathbf{\rhoa\%_p}$} &  \textbf{$\mathbf{\rhoa\%_\rho\rhoa\%_p}$} &   \textbf{$\mathbf{\rhoa\%_{\rho p}}$}  \\  \cmidrule(l){3-6}
\multicolumn{1}{|r}{\textbf{$\mathbf{\pbar}$}} & \multicolumn{1}{r|}{0.1\% } & 85.7\% & 46.6\% &39.9\% &37.9\% \\
\multicolumn{1}{|r}{\textbf{$\mathbf{}$}} & \multicolumn{1}{r|}{1.0\% } & 82.2\% & 68.3\% &56.1\% &55.2\% \\
\multicolumn{1}{|r}{\textbf{$\mathbf{}$}} & \multicolumn{1}{r|}{5.0\% } & 80.1\% & 84.2\% &67.4\% &67.1\% \\
\multicolumn{1}{|r}{\textbf{$\mathbf{}$}} & \multicolumn{1}{r|}{20.0\% } & 78.6\% & 91.9\% &72.2\% &72.2\% \\
\multicolumn{1}{|r}{\textbf{$\mathbf{}$}} & \multicolumn{1}{r|}{50.0\% } & 78.1\% & 93.7\% &73.2\% &73.1\% \\
\end{tabular}}
\caption{Measured asset correlation as percentage of actual mean asset correlation, using $n=10^9$, $K=200$, $L=100$ , $ \rhoabar =4\% $, $\tau=0\%$, $s(\rhoa)=20\%$, $s(p)=20\%$.}\label{table_zero1}
\end{table}
\begin{table}[htbp]
\centering \footnotesize{
\begin{tabular}{@{}llllll@{}}
\cmidrule(l){3-6} 
\textbf{} & \textbf{} & \textbf{$\mathbf{\rhoa\%_\rho}$} &  \textbf{$\mathbf{\rhoa\%_p}$} &  \textbf{$\mathbf{\rhoa\%_\rho\rhoa\%_p}$} &   \textbf{$\mathbf{\rhoa\%_{\rho p}}$}  \\  \cmidrule(l){3-6}
\multicolumn{1}{|r}{\textbf{$\mathbf{\pbar}$}} & \multicolumn{1}{r|}{0.1\% } & 101.6\% & 45.2\% &45.9\% &44.5\% \\
\multicolumn{1}{|r}{\textbf{$\mathbf{}$}} & \multicolumn{1}{r|}{1.0\% } & 99.1\% & 67.6\% &67.0\% &66.3\% \\
\multicolumn{1}{|r}{\textbf{$\mathbf{}$}} & \multicolumn{1}{r|}{5.0\% } & 97.6\% & 84.1\% &82.0\% &81.8\% \\
\multicolumn{1}{|r}{\textbf{$\mathbf{}$}} & \multicolumn{1}{r|}{20.0\% } & 96.4\% & 91.9\% &88.7\% &88.6\% \\
\multicolumn{1}{|r}{\textbf{$\mathbf{}$}} & \multicolumn{1}{r|}{50.0\% } & 96.0\% & 93.7\% &90.0\% &90.0\% \\
\end{tabular}}
\caption{Measured asset correlation as percentage of actual mean asset correlation, using $n=10^9$, $K=200$, $L=100$ , $ \rhoabar =20\% $, $\tau=0\%$, $s(\rhoa)=20\%$, $s(p)=20\%$.}\label{table_zero2}
\end{table}

We have checked the validity of the relationship (\ref{stacking}) for all combinations of $\rhoabar$, $\pbar$, $\sigma (p)$, $\sigma (\rho )$ with $\rhoabar \in\{ 1\% , 4\% , 12\% , 20\% , 40\%\}$, $\pbar \in\{ 0.1\% , 1\% , 2\% , 5\% , 10\%\ , 20\%\}$, $\sigma(p) \in\{ 5\% , 20\% , 50\% \}$, $\sigma (\rho) \in\{ 5\% , 20\% , 50\% \}$ and found it to hold always. In fact for every combination studied $ \rhoa\%_{\rho p} \le \rhoa\%_\rho \rhoa\%_p$ was true, i.e. the stacking effect was never weaker than multiplicative.

\begin{figure}[!htbp]
\centering
\includegraphics[width=0.8\linewidth]{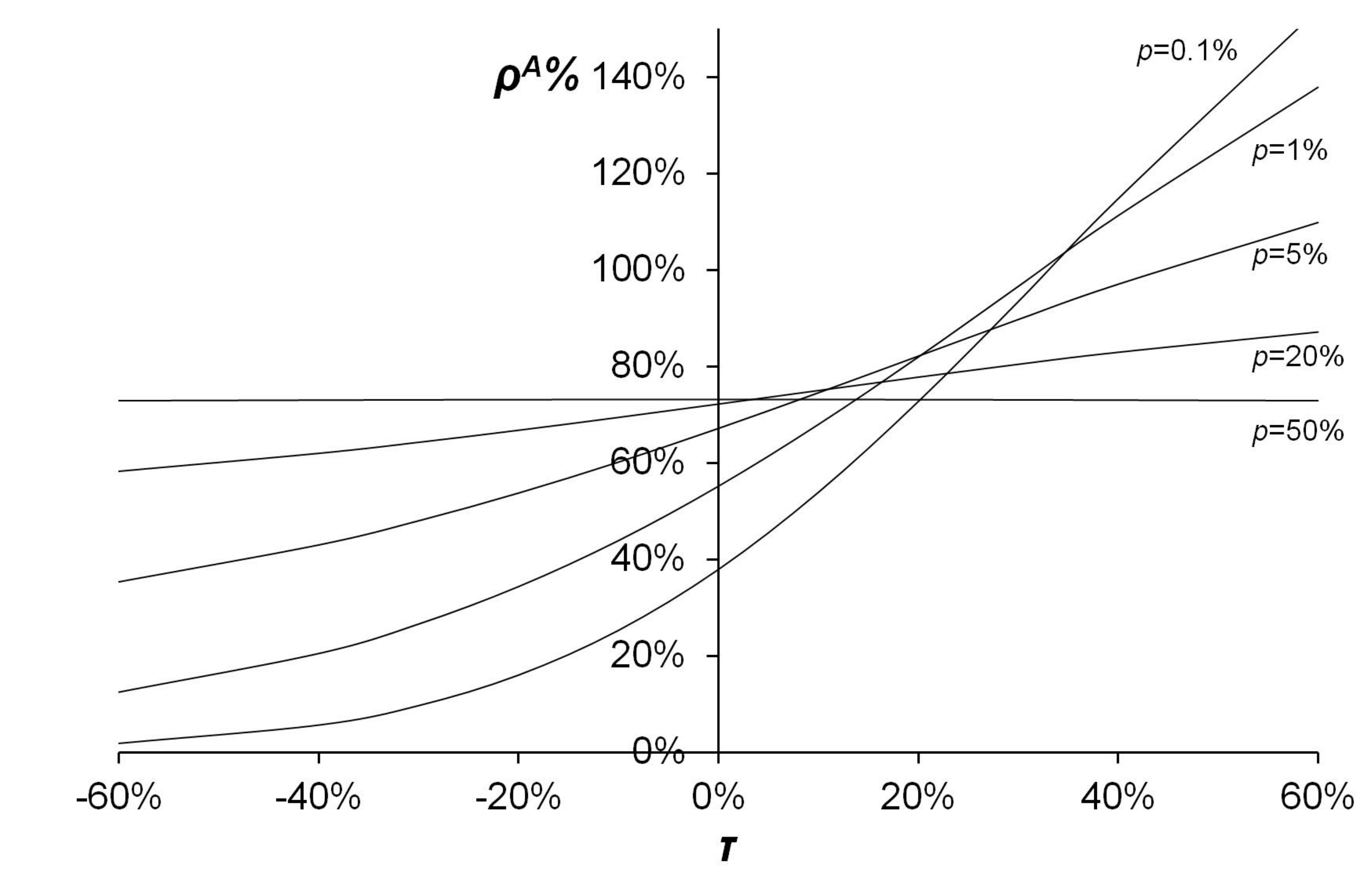}
\caption{Measured asset correlation as percentage of actual mean asset correlation, using $n=10^9$, $K=200$, $L=100$ , $ \rhoabar =4\%$, $s(\rhoa)= 20\%$, $ s(p) = 20\%$.  }
 \label{figure_onion_13}
\end{figure}

\begin{figure}[!htbp]
\centering
\includegraphics[width=0.8\linewidth]{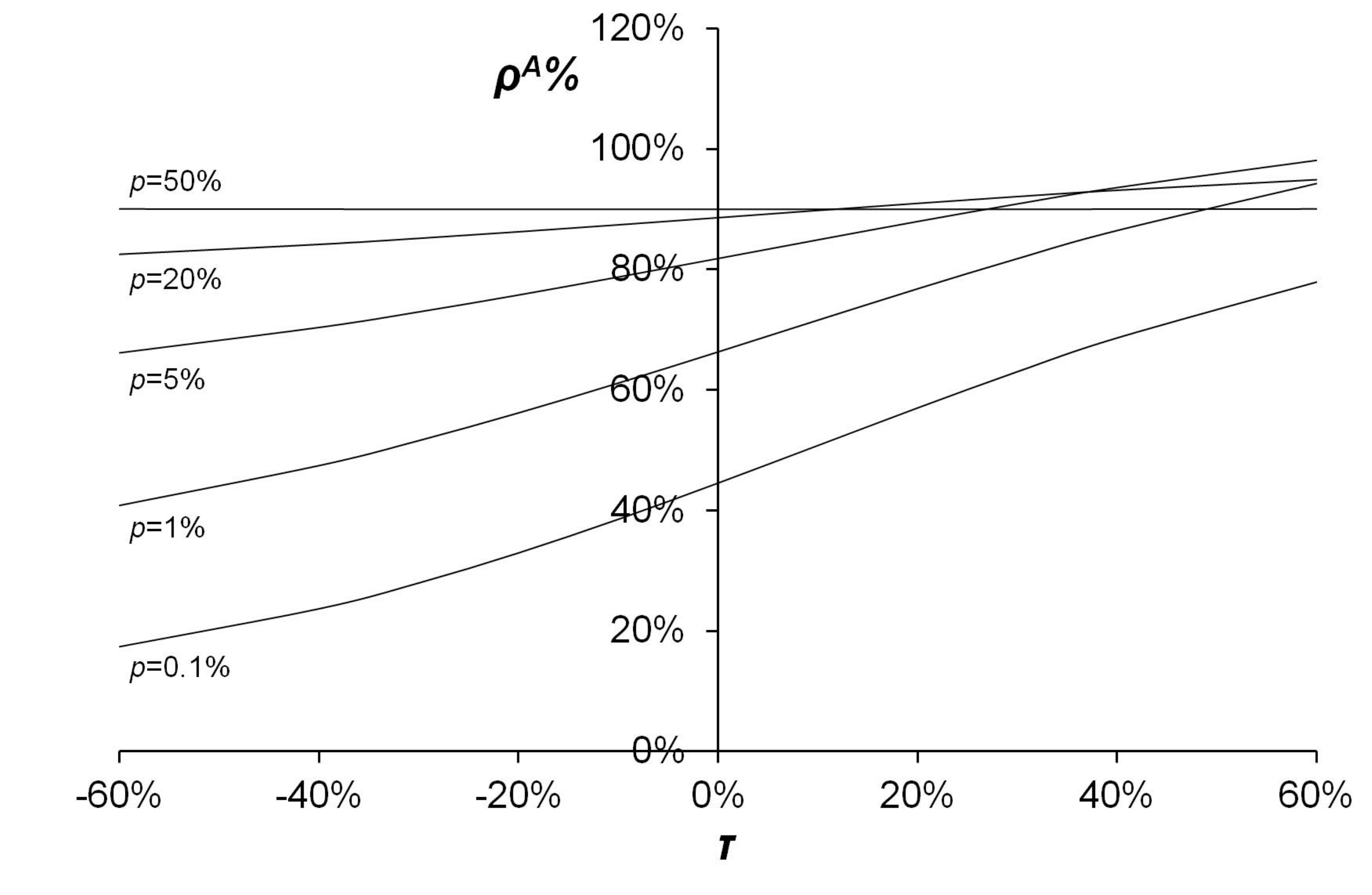}
\caption{Measured asset correlation as percentage of actual mean asset correlation, using $n=10^9$, $K=200$, $L=100$ , $ \rhoabar =20\%$, $s(\rhoa)= 20\%$, $ s(p) = 20\%$.  }
 \label{figure_onion_14}
\end{figure}

The figures \ref{figure_onion_13} and \ref{figure_onion_14} show for the same choices of $\rhoabar$ and $\pbar$ as used for the tables \ref{table_zero1} and \ref{table_zero2}, how a negative correlation between $p$ and $\rhoa$ can increase the asset correlation reduction even further and how a positive correlation can mitigate it. For $\pbar=50\%$ no effect of the correlation on the outcome of results was observed.

\FloatBarrier

%% file: paper_literature.tex
The inhomogeneity effect theoretically analyzed in this study can be observed in practice. \cite{Dietsch_et_al_2004} report asset correlation estimates for French SMEs differentiated by rating grade, see table \ref{table_dietsch}: the asset correlation estimate based on the whole pool is lower than all the estimates that are restricted to only one rating grade bucket. Clearly the inhomogeneity with respect to the PD of the whole exposure pool is stronger than the inhomogeneity of any one rating grade bucket taken separately, such that this data is consistent with the findings of the present study.

\begin{table}[!htbp]
\centering
\begin{minipage}[b]{.4\textwidth}
  \centering
\footnotesize{
\begin{tabular}{ll}
\cmidrule(l){1-2} 
\textbf{Rating} & \textbf{$\mathbf{\rhoa}$}   \\  \cmidrule(l){1-2}
1 (high) & 0.0219\\
2 & 0.0229 \\
3 & 0.0231 \\
4 & 0.0267 \\
5 & 0.0151 \\
6 & 0.0199 \\
7 & 0.0298 \\
8 (low) & 0.0307 \\ \cmidrule(l){1-2}
Total & 0.0128 \\ \cmidrule(l){1-2}
\end{tabular}}
\caption{Measured asset correlations for French SMEs reported by {\protect\cite{Dietsch_et_al_2004}}, cited in {\protect\cite{Hashimoto_2009}}.}\label{table_dietsch}
\end{minipage}
\qquad
\begin{minipage}[b]{.4\textwidth}
  \centering
  \footnotesize{
\begin{tabular}{ll}
\cmidrule(l){1-2} 
\textbf{Rating} & \textbf{$\mathbf{\rhoa}$}   \\  \cmidrule(l){1-2}
BB & 0.06042 \\
B & 0.04516 \\
CCC & 0.06949 \\ \cmidrule(l){1-2}
All grades & 0.03948\\ \cmidrule(l){1-2}
\end{tabular}}
\caption{Measured asset correlations for corporates, based on Standard
\& Poor’s data, adapted from {\protect\cite{Hamerle_et_al_2003}}.}\label{table_hamerle}
\end{minipage}
\end{table}

\cite{Hamerle_et_al_2003} (table 1) use Standard \& Poor’s ratings data spanning 1982 to 1999 to estimate asset correlations for corporates. If they assume constant PD over time (an assumption also made in this paper) then they report the results given in table \ref{table_hamerle}. Again, these results are consistent with the findings of the present study.

\cite{Haddad_2013} (table 5, last row) report a similar picture for Canadian SMEs and \cite{Castro_2012} (tables 6, 7, 9) shows results for Moody\rq{}s rating data from 1970 to 2009 spanning different geographies and  industries. The asset correlations estimated via the whole pool is always lower than the asset correlation estimated in a particular region or for a particular industry, independent of the modeling assumption used for the estimation. 

\cite{Demey_et_al_2004} perform Monte Carlo experiments to study the small sample properties of various asset correlation estimators. One aspect they studied was the effect of an exposure pool that is inhomogeneous with respect to the PD. The results they present (tables A, B and C in \cite{Demey_et_al_2004}) show reduced asset correlation estimates for the inhomogeneous exposure pool. \cite{Demey_et_al_2004} also derive asset correlation estimates for different industry sectors from S\&P default data and observe that if they estimate asset correlations for all industry sectors combined then the estimated asset correlations are reduced (c.f. remark 1 in \cite{Demey_et_al_2004}).

\cite{Kalkbrener_et_al_2009} use Standard \& Poor’s ratings data spanning 1981 to 2009 to calculate asset correlations for 13 industry segments using different maximum likelihood estimators. In the first column of their table 2 they show the asset correlation for each industry segment estimated without taking rating information into account, i.e. implicit\-ly assuming homogeneity with respect to the PD. In the second column of their table 2 they show asset correlations for each industry estimated under the assumption that each industry segment is sub-divided in 7 rating classes and homogeneity with respect to the PD is only assumed within those rating classes. Comparing the results they find that in all but one industry segment the correlation estimate in the first column is lower than the estimate in the second column. Averaged across all industry segments the estimated correlation in the first column is $16.3\%$ versus $19.8\%$ in the second column. 

Not all observations reported in the literature give such a clear picture as the observations described so far, where the asset correlation measured in a pool of exposures is smaller than the asset correlation measured in every part of that pool. But for being consistent with the findings of this study such a clear picture is not necessary, in particular if the asset correlation estimates vary strongly across the different sub-pools. \cite{Dietsch_et_al_2004} for instance report asset correlation estimates for German SMEs, where it is visible that overall asset correlation estimates tend to be lower than estimates calculated for one PD bucket alone, but there are some outliers. \cite{Duellmann_et_al_2013} provide a comprehensive overview of asset correlation studies some of which give further evidence of the inhomogeneity effect described in this paper.

%% file: paper_discussion.tex
The study presented in this paper shows that in the cases most relevant to the practitioner the asset correlation is underestimated if it is estimated from the default rate time series of an inhomogeneous exposure pool. This underestimation is separate from the estimation bias that is downward for most estimators in use today (see e.g. \cite{Duellmann_et_al_2008}) and that typically is reduced if the default time series becomes longer (c.f. \cite{Gordy_et_al_2010}). It is left to a subsequent paper to study the way how the inhomogeneity effect combines with the estimation bias of different asset correlation estimators. What can be said, however, is that in the extreme case of $s(p)=1$ the observed default rate will be constant over time and any useful asset correlation estimator  $\rhoahat$ will give $\rhoahat\approx 0$ as a consequence independent of $\rhoabar$ and the length of the default time series, so at least in this extreme case no realistic estimator will be able to compensate for the inhomogeneity effect.

Even if the asset correlation is estimated from one PD bucket alone one has to expect an underestimation because the rating system that has assigned an identical PD to all exposures in the PD bucket will not have taken into account all information available and the true PDs that would have been assigned by a `perfect' rating system will be inhomogeneous. How relevant this inhomogeneity can be has not been studied, but most likely it will lead to an underestimation of the asset correlation and this effect will be the larger the lower the discriminatory power of the rating system assigning the PDs has been. In a similar fashion one will always have inhomogeneous asset correlations in the exposure pool considered, which also leads to an underestimation of asset correlations in many circumstances, as we have seen in section \ref{inh2}. If asset correlations are directly measured from time series of asset value changes then this source of underestimation does not exist and hence the findings of this paper provide an explanation for observation that asset correlations estimated from default time series tend to be lower than asset correlations measured from asset value data (c.f.  \cite{Duellmann_et_al_2008}, \cite{Frye_2008},  \cite{Kalkbrener_et_al_2009} and \cite{Chernih_et_al_2010}). 

\begin{figure}[!htb]
\centering
\begin{subfigure}{.5\textwidth}
  \centering
  \includegraphics[width=.95\linewidth]{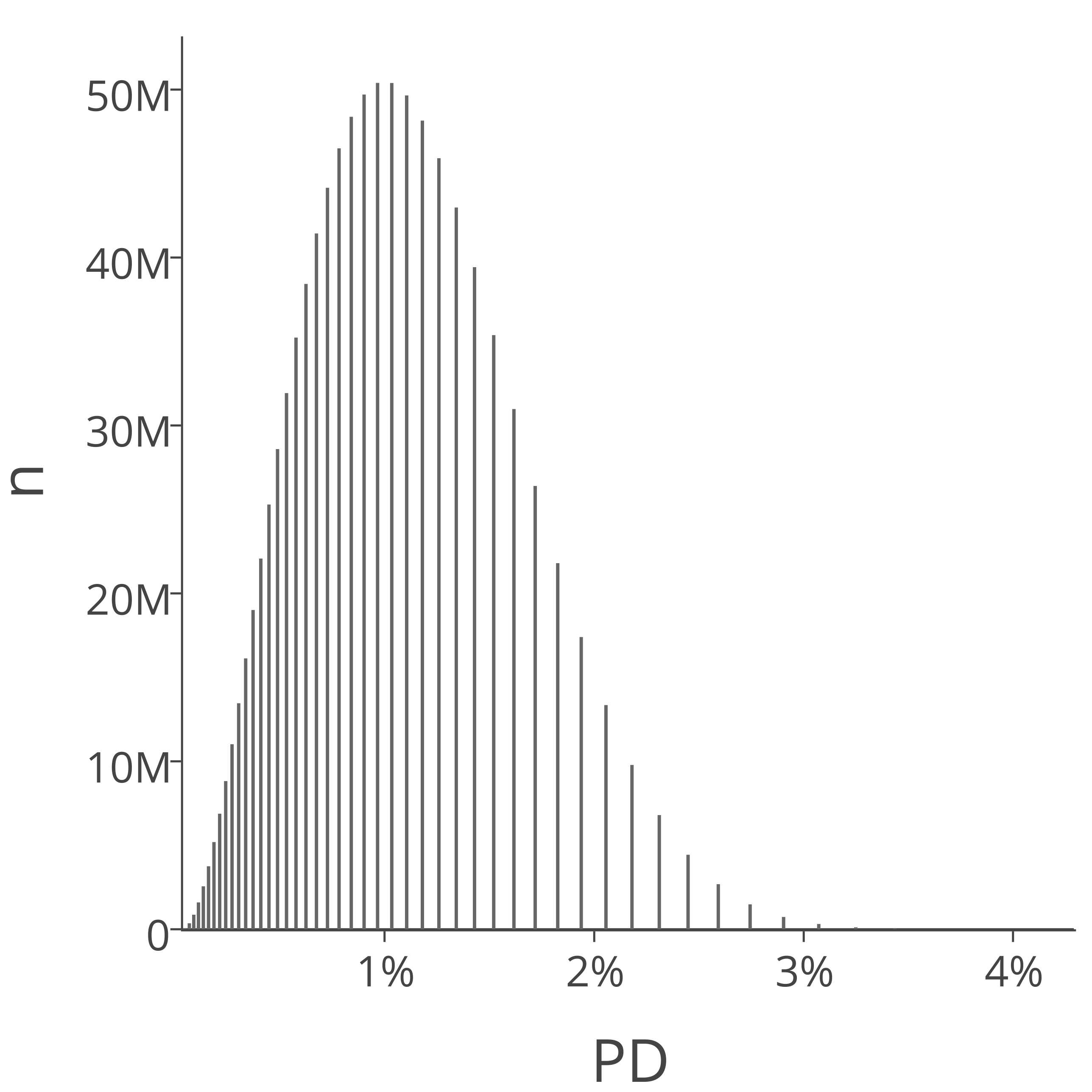}
\end{subfigure}%
\begin{subfigure}{.5\textwidth}
  \centering
  \includegraphics[width=.95\linewidth]{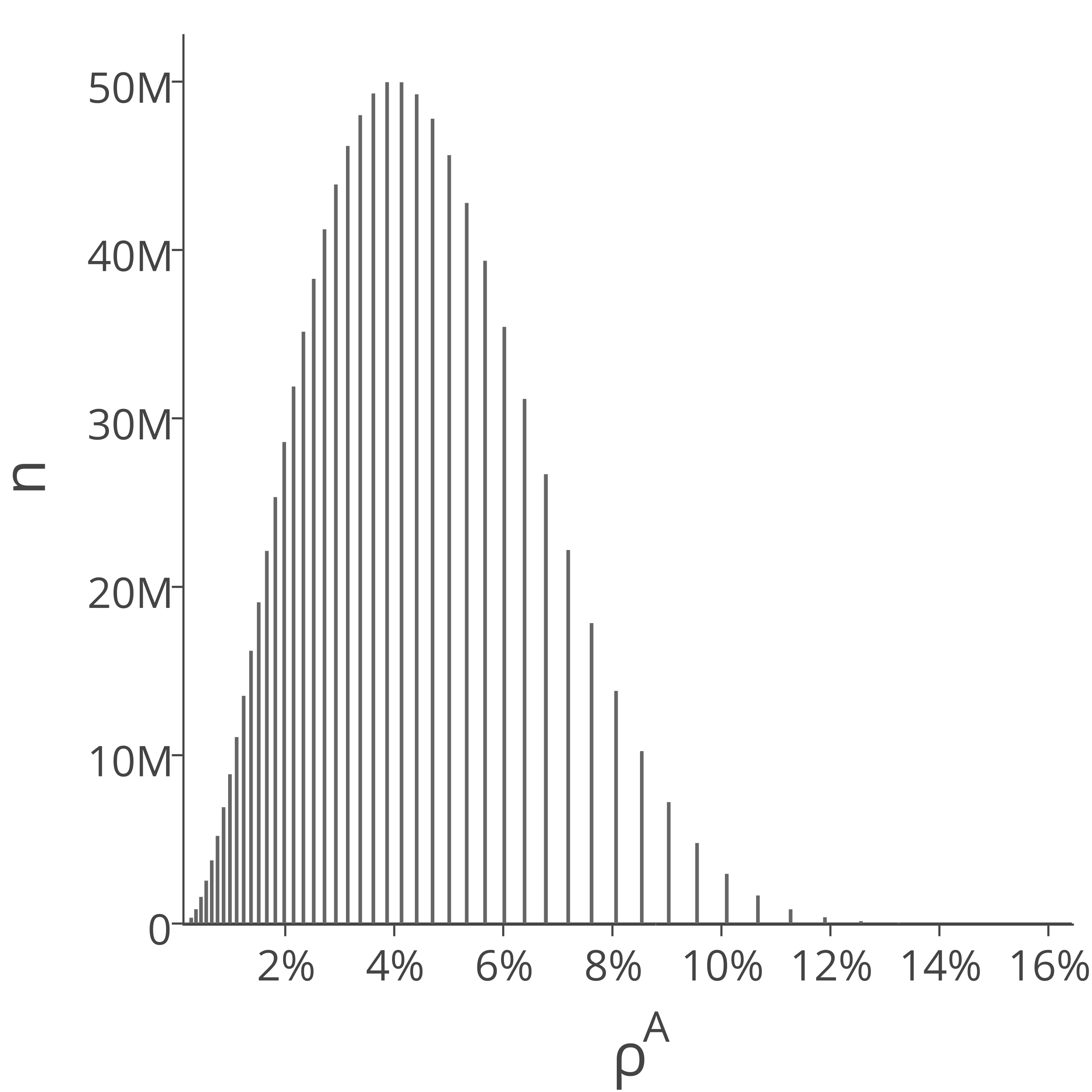}
\end{subfigure}
\begin{subfigure}{1.0\textwidth}
\centering
  \includegraphics[width=.95\linewidth]{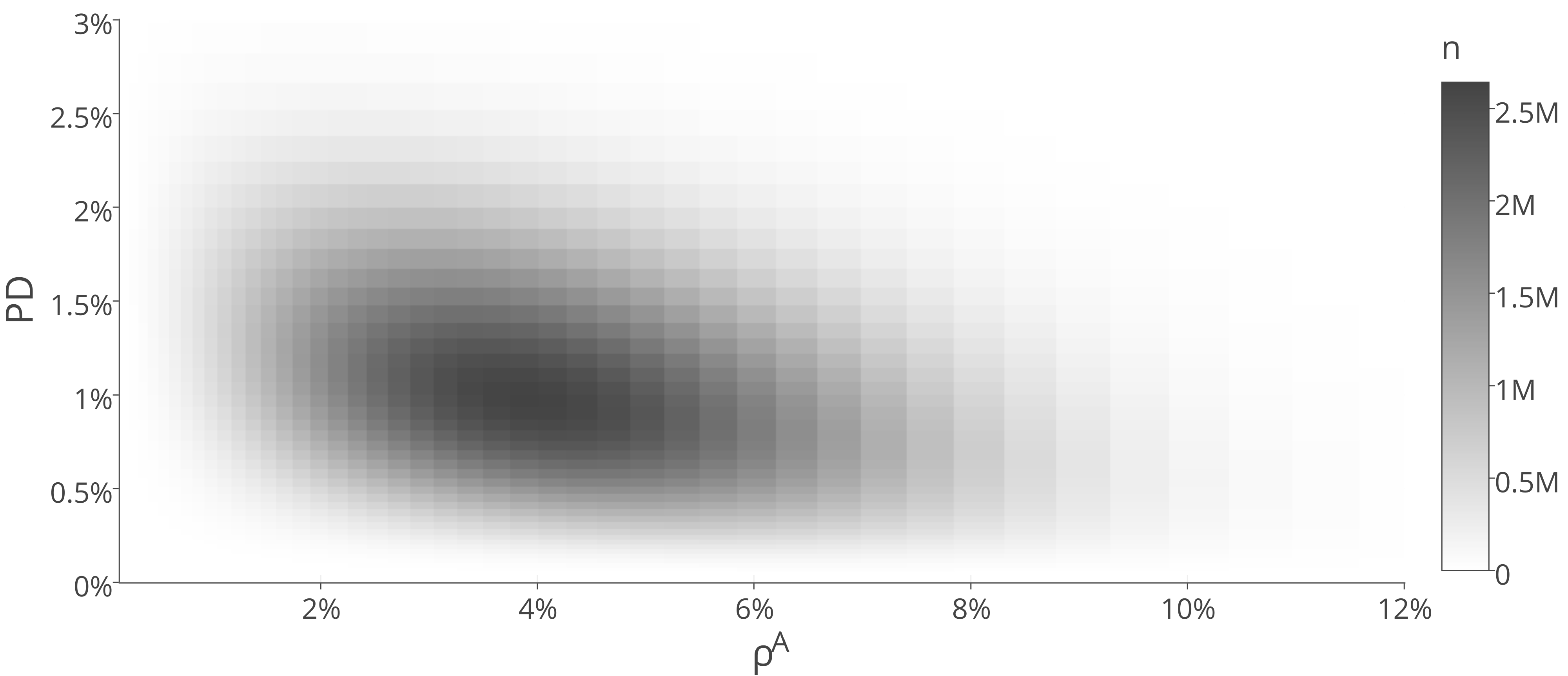}
\end{subfigure}
  \caption{exposure constellation for $n=10^9$, $K=50$, $L=50$, $\pbar=1\%$, $s(p)=5\%$,  $\rhoabar=4\%$, $s(\rhoa)=10\%$, $\tau=-20\%$, $\rhoa\%=85.8\%$}\label{example2}
\end{figure}

The big reductions in asset correlation described in section \ref{inh1} and depicted in figure \ref{figure_onion_02} require a rather strong inhomogeneity with respect to the PD and one might argue that with some care such a strong inhomogeneity can be avoided in practice. However, the observation in section \ref{inh3} that the effects of the inhomogeneities with respect to PD and asset correlation stack up and can be further increased by a negative correlation between PD and asset correlation leads to the situation that already rather limited inhomogeneities together with some moderately negative correlation lead to a noticeable reduction of the measured asset correlation. For instance, if we choose $\pbar=1\%$, $\sigma(p)=0.5\%$ (and hence $s(p)=5\%$) as well as $\rhoabar = 4\%$, $\sigma(\rhoa)=2\%$ (and hence $s(\rhoa)=10\%$) and $\tau=-20\%$, then the observed reduction of the asset correlation is already close to $15\%$. Figure \ref{example2} shows the marginal PD- and $\rhoa$-distributions as well as the joint PD-$\rhoa$-distribution for the corresponding exposure constellation.

None of these assumptions is unrealistic if the PD is measured with a rating system that by its nature cannot incorporate all predictive information and if the homogeneity of the asset correlations in a given risk segment is only ensured by the assignation of exposures to that risk segment based on characteristics such as industry and region. Note that there is some evidence that larger companies tend to have a higher asset correlation (because they depend less on idiosyncratic factors to drive their risk) as well as a lower PD, such that a negative correlation between PD and asset correlation is induced if the exposure pool is not homogeneous with respect to company size \cite{Lee_et_al_2009}. The Basel RWA formula implicitly assumes a negative correlation between asset correlation and PD. Refer also to \cite{Chernih_et_al_2010} (figure 1) and \cite{Tarashev_2007} (table 3) for some empirical evidence as well as to \cite{Lopez_2004} and \cite{Lee_et_al_2009} for a more in depth discussion of this topic.

While we believe that the effects described in this paper are real and important, it is worth highlighting the assumptions we have made coming to the conclusions presented: we have assumed that the structural model of default and the factor model presented in section \ref{theo} are in fact applicable and that the exposure constellation remains constant over time. We also made the assumption that the PD and asset correlation profiles can be approximated by a beta distribution and that their dependence can be modeled by a Gaussian copula. Those distributional assumptions clearly influence the numerical results reported in this paper but arguably not its main conclusion that neglecting the inhomogeneity effect leads to reduced asset correlation measurements.

%% file: paper_appendix.tex
Given a certain input configuration $\{n, \pbar , \sigma (p), \rhoabar , \sigma (\rhoa), \tau\}$ we need to construct an exposure constellation that corresponds to it. 

The chosen approach is to have beta distributions across the PD dimension as well as the $\rhoa$ dimension and a Gaussian copula for the dependence between $\rhoa$ and the PD. The standard beta distribution has support $[0,1]$, which would be suitable for the PD and $\rhoa$ distributions, but especially for PD distributions there will often be hardly any mass close to PD=1, such that the support of the beta distributions can be reduced for numerical efficiency. Note that the parameters $\alpha$ and $\beta$ of the beta distribution can be chosen such that it takes any mean $\mu\in ]0,1[$ and any variance $0< \sigma^2<\mu(1-\mu)$ (c.f. \cite{Tasche_2016}, appendix 1).

If $K=1$ we simply set $p_1=\pbar$. If $K>1$ we determine $p_{\mathrm{min}}$ and $p_{\mathrm{max}}$  as follows:
\begin{eqnarray}
 p_{\mathrm{min}}&=&F^{-1}_{\pbar,\sigma(p)}\left(\frac{1}{Kg}\right), \quad p_{\mathrm{max}}=F^{-1}_{\pbar,\sigma(p)}\left(1-\frac{1}{Kg}\right)\nonumber
\end{eqnarray}
Here $F^{-1}_{\pbar,\sigma(p)}$ stands for the inverse beta distribution function with mean $\pbar$, standard deviation $\sigma(p)$ and support $[0,1]$; $g$ is a factor that determines how strongly the support of the beta distribution should be reduced. It has little effect on results unless very extreme PD distributions are considered or an extreme $\tau$. Unless otherwise stated $g=1000$ has been used.

It is suboptimal numerically to divide the range $[ p_{\mathrm{min}},p_{\mathrm{max}}]$ into equal sized buckets, as the PD profile often is very skewed. We therefore define a midpoint  $p_{\mathrm{mid}}$ and ensure that  both $[ p_{\mathrm{min}},p_{\mathrm{mid}}]$ and  $[ p_{\mathrm{mid}},p_{\mathrm{max}}]$ are divided into $K / 2$ buckets. As midpoint we choose $ p_{\mathrm{mid}}=\pbar$ unless otherwise stated, other options are e.g. $ p_{\mathrm{mid}}=p_{\mathrm{mode}}$ or $ p_{\mathrm{mid}}=p_{\mathrm{median}}$. We define
\begin{equation}
t=\frac{p_{\mathrm{mid}}-p_{\mathrm{min}}}{p_{\mathrm{max}}-p_{\mathrm{min}}}\nonumber
\end{equation}
and the bucket limits $b_p(m), 0\le m\le K$
\begin{equation}
b_p(m)=\left\{\begin{array}{ll}p_{\mathrm{min}}+\left(p_{\mathrm{max}}-p_{\mathrm{min}}\right)\frac{m}{K} & \mbox{if } t=\frac{1}{2} \\ p_{\mathrm{min}}+\left(p_{\mathrm{max}}-p_{\mathrm{min}}\right)\frac{t^2}{1-2t}\left( \left( \frac{1-t}{t} \right)^\frac{2m}{K} -1 \right) & \mbox{if } t\ne\frac{1}{2}\end{array} \right.\nonumber
\end{equation}

Note that the limiting case $t=1 / 2$ corresponds to a symmetric PD distribution and an even distribution of buckets is suitable. For $t\neq 1 / 2$ the given formula achieves that the size of the PD buckets varies smoothly across the range. By construction we have
\begin{equation}
b_p(0)=p_{\mathrm{min}},\quad b_p(K)=p_{\mathrm{max}},\quad b_p(K/2)=p_{\mathrm{mid}}\nonumber
\end{equation}
and we can define the PD values assigned to each of the $K$ PD-buckets:
\begin{equation}
1\le k\le K: p_k=\frac{b_p(k)-b_p(k-1)}{2}\label{bucketv}
\end{equation}
In an analogous fashion we also define the $\rhoa$-bucket limits $b_\rho(m), 0\le m \le L$ and the $\rhoa$ values assigned to each of the $L$ $\rhoa$-buckets $\rhoa_l , 1\le l \le L$.

In order to distribute the exposures among the $L\times K$ exposure buckets we define an auxiliary variable $x_{kl}$ for $1\le k \le K, 1\le l\le L$:
\begin{equation}
x_{kl}=C\left(B_p \left(b_p (k)\right), B_\rho (b_\rho (l)), \sin\left(\frac{\pi}{2}\tau\right)\right)\label{auxi}
\end{equation}
where $C(u, v, \rho)=\Phi_2\left(\Phi^{-1}(u), \Phi^{-1}(v), \rho\right)$ is the bivariate normal copula continuously extended to $u, v = \pm 1$ and $B_p$ is the cumulative Beta distribution, with mean $\pbar$ and standard deviation $\sigma (p)$ defined on support  $[ p_{\mathrm{min}},p_{\mathrm{max}}]$. Note that typically the Beta distribution is defined on $[0,1]$, but the linear transformation to shrink the support is straightforward. Similarly  $B_\rho$ is the cumulative Beta distribution, with mean $\rhoabar$ and standard deviation $\sigma (\rhoa)$ defined on support  $[ \rhoa_{\mathrm{min}},\rhoa_{\mathrm{max}}]$.

Note that in case of a bivariate normal copula and infinitely many buckets, Kendall\rq{}s $\tau$ and the linear correlation coefficient $\rho$ are linked via  $\rho=\sin\left(\frac{\pi}{2}\tau\right)$, see e.g. \cite{Lindskog_2003}. We choose to take  Kendall\rq{}s $\tau$ as input instead of the linear correlation coefficient $\rho$, because we use non-normal marginal distributions such that the linear correlation to be seen in the resulting exposure constellation will be different to the linear correlation $\rho=\sin\left(\frac{\pi}{2}\tau\right)$ originally used in equation (\ref{auxi}). Kendall\rq{}s $\tau$  is invariant to changing marginal distributions and therefore more suitable for our study: we expect the $\tau$ seen in the resulting exposure constellation to be equal to the $\tau$ used in equation (\ref{auxi}), at least if $K$ and $L$ are large enough to limit the error due to the bucketing.

Using $x_{kl}$ the number of exposures in each bucket $n_{kl}$ can now be calculated iteratively:
\begin{equation}
n_{kl}=\max\left(0 , \left\lfloor \frac{1}{2} + nx_{kl} +\sum_{i=1}^{k-1}\sum_{j=1}^{l-1} n_{ij} -\sum_{i=1}^{k}\sum_{j=1}^{l-1} n_{ij}-\sum_{i=1}^{k-1}\sum_{j=1}^{l} n_{ij} \right\rfloor\right)  \label{iterative}
\end{equation}
Note that this iterative definition ensures that the error due to enforcing an integer $n_{kl}$ does not build up and the overall number of exposures seen in the resulting exposure constellation will be very close to $n$.

Equation (\ref{iterative}) together with equation  (\ref{bucketv}) let us calculate  an exposure  constellation $\{p_k , \rhoa_l , n_{kl}, K, L\}$ for any desired number of buckets $K$ and $L$ and any given self-consistent input configuration $\{n, \pbar , \sigma (p), \rhoabar , \sigma (\rhoa), \tau\}$. The more buckets are used and the bigger $n$ is, the better the exposure constellation $\{p_k , \rhoa_l , n_{kl}, K, L\}$ will  reflect the given input configuration $\{n, \pbar , \sigma (p), \rhoabar , \sigma (\rhoa), \tau\}$, as can be checked directly by diagnosing  $\{n, \pbar , \sigma (p), \rhoabar , \sigma (\rhoa), \tau\}$ from $\{p_k , \rhoa_l , n_{kl}, K, L\}$. Unless otherwise stated the parameters of the original input configuration and the parameters diagnosed from the exposure constellation do not differ by more than $1\%$ for the input configurations studied in section \ref{num}.

%% file: Wunderer_2019_asset_correlation_estimation_for_inhomogeneous_exposure_pools.bbl
\begin{thebibliography}{}

\bibitem[\protect\astroncite{Bluhm and Overbeck}{2007}]{Bluhm_et_al_2007}
Bluhm, C. and Overbeck, L. (2007).
\newblock {\em Structured Credit Portfolio Analysis, Baskets \& CDOs}.
\newblock CRC Press.

\bibitem[\protect\astroncite{Bluhm et~al.}{2010}]{Bluhm_et_al_2010}
Bluhm, C., Overbeck, L., and Wagner, C. (2010).
\newblock {\em An Introduction to Credit Risk Modeling, second edition}.
\newblock CRC Press.

\bibitem[\protect\astroncite{Castro}{2012}]{Castro_2012}
Castro, C. (2012).
\newblock Confidence sets for asset correlations in portfolio credit risk.
\newblock {\em Revista de Economia del Rosario}, 15(1):19--58.

\bibitem[\protect\astroncite{Chernih et~al.}{2010}]{Chernih_et_al_2010}
Chernih, A., Henrard, L., and Vanduffel, S. (2010).
\newblock Reconciling credit correlations.
\newblock {\em The Journal of Risk Model Validation}, 4(2):47--64.

\bibitem[\protect\astroncite{Demey et~al.}{2004}]{Demey_et_al_2004}
Demey, P., Jouanin, J.-F., Roget, C., and Roncalli, T. (2004).
\newblock Maximum likelihood estimate of default correlations.
\newblock {\em RISK}, pages 104--108.
\newblock An important correction can be found at
  \url{http://www.thierry-roncalli.com/download/correction-mledc.pdf}.

\bibitem[\protect\astroncite{Dietsch and Petey}{2004}]{Dietsch_et_al_2004}
Dietsch, M. and Petey, J. (2004).
\newblock Should sme exposures be treated as retail or corporate exposures? a
  comparative analysis of default probabilities and asset correlations in
  french and german smes.
\newblock {\em Journal of Banking and Finance}, 28:773--788.

\bibitem[\protect\astroncite{D{\"u}llmann and
  Koziol}{2013}]{Duellmann_et_al_2013}
D{\"u}llmann, K. and Koziol, P. (2013).
\newblock Evaluation of minimum capital requirements for bank loans to smes.
\newblock Technical Report 22/2013, Deutsche Bundesbank.

\bibitem[\protect\astroncite{D{\"u}llmann et~al.}{2008}]{Duellmann_et_al_2008}
D{\"u}llmann, K., K{\"u}ll, J., and Kunisch, M. (2008).
\newblock Estimating asset correlations from stock prices or default rates –
  which method is superior?
\newblock Technical Report 4/2008, Deutsche Bundesbank.

\bibitem[\protect\astroncite{Frei and Wunsch}{2018}]{Frei_et_al_2018}
Frei, C. and Wunsch, M. (2018).
\newblock Moment estimators for autocorrelated time series and their
  application to default correlations.
\newblock {\em Journal of Credit Risk}, 14(1):1--29.

\bibitem[\protect\astroncite{Frye}{2008}]{Frye_2008}
Frye, J. (2008).
\newblock Correlation and asset correlation in the structural portfolio model.
\newblock {\em Journal of credit risk}, 4(2).

\bibitem[\protect\astroncite{Gordy and Heitfield}{2010}]{Gordy_et_al_2010}
Gordy, M. and Heitfield, E. (2010).
\newblock Small-sample estimation of models of portfolio credit risk.
\newblock In {\em Recent Advances in Financial Engineering: The Proceedings of
  the KIER-TMU International Workshop on Financial Engineering 2009}. World
  Scientific Publishing.

\bibitem[\protect\astroncite{Haddad}{2013}]{Haddad_2013}
Haddad, J.~M. (2013).
\newblock Comprehensive correlation and capital estimates for a canadian sme
  portfolio: Implications for basel.
\newblock {\em unpublished working paper}.
\newblock Available at
  \url{http://www.greta.it/credit/credit2013/PAPERS/Speaker/thursday/11_Haddad.pdf}.

\bibitem[\protect\astroncite{Hamerle et~al.}{2003}]{Hamerle_et_al_2003}
Hamerle, A., Liebig, T., and R{\"o}sch, D. (2003).
\newblock Credit risk factor modeling and the basel {II} {IRB} approach.
\newblock Technical Report 02/2003, Deutsche Bundesbank.

\bibitem[\protect\astroncite{Hashimoto}{2009}]{Hashimoto_2009}
Hashimoto, T. (2009).
\newblock Asset correlation for credit risk analysis - empirical study of
  default data for japanese companies -.
\newblock Technical Report 09-E-3, Bank of Japan.

\bibitem[\protect\astroncite{Kalkbrener and
  Onwunta}{2009}]{Kalkbrener_et_al_2009}
Kalkbrener, M. and Onwunta, A. (2009).
\newblock Validating structural credit portfolio models.
\newblock {\em COMISEF Working Papers Series}, (WPS-014).
\newblock Available at \url{http://comisef.eu/files/wps014.pdf}.

\bibitem[\protect\astroncite{Lee et~al.}{2009}]{Lee_et_al_2009}
Lee, J., Wang, J., and Zhang, J. (2009).
\newblock The relationship between average asset correlation and default
  probability.
\newblock Technical report, Moody\rq{}s KMV.

\bibitem[\protect\astroncite{Lindskog et~al.}{2003}]{Lindskog_2003}
Lindskog, F., McNeil, A., and Schmock, U. (2003).
\newblock Kendall’s tau for elliptical distributions.
\newblock In {\em Credit Risk: Measurement, Evaluation and Management,
  Physica-Verlag}, pages 149--156. Physica.

\bibitem[\protect\astroncite{Lopez}{2004}]{Lopez_2004}
Lopez, J.~A. (2004).
\newblock The empirical relationship between average asset correlation, firm
  probability of default, and asset size.
\newblock {\em Journal of Financial Intermediation}, 13(2):265--283.

\bibitem[\protect\astroncite{Lucas}{1995}]{Lucas_1995}
Lucas, D.~J. (1995).
\newblock Default correlation and credit analysis.
\newblock {\em The journal of fixed income}, pages 76--87.

\bibitem[\protect\astroncite{McNeil et~al.}{2015}]{McNeil_et_al_2015}
McNeil, A.~J., Frey, R., and Embrechts, P. (2015).
\newblock {\em Quantitative Risk Management}.
\newblock Princeton University Press.

\bibitem[\protect\astroncite{Meyer}{2009}]{Meyer_2009}
Meyer, C. (2009).
\newblock Estimation of intra-sector asset correlations.
\newblock {\em The Journal of Risk Model Validation}, 3(3):47--79.

\bibitem[\protect\astroncite{SAS}{1999}]{SAS_1999}
SAS (1999).
\newblock Measures of association, {SAS} {O}nline{D}oc®.
\newblock Available at
  \url{http://v8doc.sas.com/sashtml/stat/chap28/sect20.htm}.

\bibitem[\protect\astroncite{Tarashev and Zhu}{2007}]{Tarashev_2007}
Tarashev, N. and Zhu, H. (2007).
\newblock Modelling and calibration errors in measures of portfolio credit
  risk.
\newblock {\em BIS Working Papers}, 230.

\bibitem[\protect\astroncite{Tasche}{2016}]{Tasche_2016}
Tasche, D. (2016).
\newblock Fitting a distribution to value-at-risk and expected shortfall, with
  an application to covered bonds.
\newblock {\em Journal of Credit Risk}, 12(2):77--111.

\end{thebibliography}
